\documentclass[%
 reprint,
 superscriptaddress,
 amsmath,amssymb,
 aps,
 prc,
 floatfix
]{revtex4-2}

\usepackage{CJK}
\usepackage{graphicx}
\usepackage[colorlinks=true,linkcolor=blue,citecolor=blue,urlcolor=blue]{hyperref}

\begin{document}
\begin{CJK*}{UTF8}{gbsn}

\title{Light-particle emission in nucleon-induced fission of \texorpdfstring{$^{235,238}\mathrm{U}$}{235,238U} at energies from \texorpdfstring{$10~\mathrm{MeV}$}{10 MeV} to \texorpdfstring{$1~\mathrm{GeV}$}{1 GeV}}

\author{Fuchang Gu (顾富昌)}
\affiliation{
Sino-French Institute of Nuclear Engineering and Technology,
Sun Yat-sen University, Zhuhai 519082, China
}
\author{Yingge Huang (黄英格)}
\affiliation{
Institute of Theoretical Physics,
Chinese Academy of Sciences, Beijing 100190, China
}
\author{Qiafeng Chen (陈洽锋)}
\affiliation{
Sino-French Institute of Nuclear Engineering and Technology,
Sun Yat-sen University, Zhuhai 519082, China
}
\author{Jun Su (苏军)}
\email[Contact author: ]{sujun3@mail.sysu.edu.cn}
\affiliation{
Sino-French Institute of Nuclear Engineering and Technology,
Sun Yat-sen University, Zhuhai 519082, China
}
\affiliation{
Key Laboratory of Nuclear Data, China Institute of Atomic Energy, Beijing 102413, China
}

\date{\today}

\begin{abstract}
\begin{description}
\setlength{\rightskip}{\leftskip}
\item[Background]
In nucleon-induced fission up to GeV energies, early nonequilibrium emission alters the residual nuclei before statistical evaporation and scission.
The emitted light particles therefore probe each reaction stage while supplying needed nuclear data.
\item[Purpose]
We study neutron and light charged-particle emission in $^{235,238}\mathrm{U}(n,f)$ and $^{235,238}\mathrm{U}(p,f)$ reactions at incident energies from $10~\mathrm{MeV}$ to $1~\mathrm{GeV}$.
\item[Methods]
The reactions are simulated event by event by coupling the isospin-dependent quantum molecular dynamics model with the GEMINI++ statistical decay code.
\item[Results]
The calculations follow the measured rise in average neutron multiplicity with incident energy, while yielding higher values above $100~\mathrm{MeV}$ and in the proton-induced prescission component.
Within the coupled calculations, the neutron multiplicity growth comes mainly from statistical evaporation before scission, whereas emission from the fission fragments changes weakly.
Although the average preequilibrium neutron yield remains small, early emission spreads the residual nuclei across multiple elements and wide excitation energies.
At high energies, protons are emitted predominantly before equilibration, directly reducing the charge retained by the residual nuclei.
\item[Conclusions]
In the coupled calculations, statistical evaporation before scission governs the neutron multiplicity growth, whereas early nonequilibrium emission shapes the residual nuclei and dominates high-energy proton emission.
These results provide a reference for modeling intermediate-energy fission, in which prescission emission changes the charge, mass, and excitation energy of the fissioning system.
\end{description}
\end{abstract}

\maketitle
\end{CJK*}

\section{\label{sec_intro}Introduction}

Nuclear fission tests our understanding of large-amplitude collective nuclear motion and supplies neutrons for nuclear energy applications~\cite{Bulgac2016,SchmidtJurado2018,Bender2020,Schunck2022}.
Applications involving energetic nucleons, including accelerator-driven subcritical systems, require fission data over a broad range of incident energies~\cite{LoMeo2015,Wang2024}.
As the incident energy rises into the intermediate-energy regime, early nonequilibrium emission processes redistribute the available energy and change the nuclei that eventually undergo fission~\cite{David2015}.
Neutron multiplicities consequently serve both as essential nuclear data and as probes of reaction evolution, with their distributions and correlations carrying information beyond the mean~\cite{Talou2018,Marini2022}.
Source-resolved light-particle emission therefore links these broad-energy data needs to the intermediate-energy fission mechanism.

In fission induced by thermal neutrons, capture forms a specific compound nucleus whose decay involves competition between fission and other channels.
At higher incident energies, early nonequilibrium emission produces residual nuclei with different masses, charges, and excitation energies~\cite{Mancusi2010PRC}.
Each residual nucleus can then evaporate particles before fissioning at a later chance, so that multichance fission proceeds from a broad distribution of residual nuclei.
Measurements of uranium fission establish a rising total neutron multiplicity over a wide energy range~\cite{Howe1984,Ethvignot2005,Mauss2026}.
Coincidence spectra and angular correlations further distinguish prescission neutrons emitted by the fissioning nucleus before scission from postscission neutrons emitted by the accelerated fragments~\cite{Cheifetz1970,Strecker1990,Fraenkel1990,*Fraenkel1991E,Batenkov2008}.
In $^{238}\mathrm{U}(p,f)$ from $20$ to $60~\mathrm{MeV}$, for example, the measured prescission neutron yield more than triples while the postscission yield stays near $4.5\text{--}5.6$~\cite{Rubchenya2001}.
Measurements of charged particles in uranium fission and of hydrogen isotopes in proton-induced lead fission probe emission that lowers the charge of the fissioning nucleus~\cite{Ivanova1957,RodriguezSanchez2016}.
These measurements, however, involve different reaction systems and particle selections.

A theoretical description must therefore pass the full ensemble of excited residual nuclei from early emission to the evaporation-fission stage, because each residual nucleus has distinct decay probabilities.
Subsequent evaporation cools the deforming nucleus and alters its mass and charge, while charged-particle branches can also populate lower-$Z$ fissioning elements~\cite{Tudora2004}.
Dynamical fission approaches, such as Langevin calculations coupled with neutron evaporation, account for the changing nucleus during collective evolution and set the scission conditions for fragment decay~\cite{Huang2022PRC}.
Three-stage descriptions connect preequilibrium emission, prescission evaporation, and fragment decay~\cite{Rubchenya2007PRC}, and microscopic calculations also track nucleon emission before and after scission~\cite{Vonta2015PRC}.
Event generators for fragment decay further describe particle multiplicity distributions beyond the means~\cite{Talou2018}.
A key question is how early nonequilibrium emission and subsequent statistical decay share the light-particle yields across particle species, projectiles, and incident energies.

Here we use the isospin-dependent quantum molecular dynamics (IQMD) model~\cite{Aichelin1991,Hartnack1998}, coupled with the statistical decay code GEMINI++~\cite{Charity2010}, following our previous work~\cite{Su2011PRC,Su2018PRC,Su2019,Song2022CPC}.
Our earlier Langevin calculations coupled with neutron evaporation identified broader preequilibrium initial conditions and charged-particle channels as directions for extension~\cite{Huang2022PRC}.
The present calculations investigate the stage-resolved light-particle multiplicities and their distributions in $^{235,238}\mathrm{U}(n,f)$ and $^{235,238}\mathrm{U}(p,f)$ at incident energies from $10~\mathrm{MeV}$ to $1~\mathrm{GeV}$.
Section~\ref{sec_model} describes the model, Sec.~\ref{sec_results} presents the results and discussion, and Sec.~\ref{sec_summary} summarizes the work.

\section{\label{sec_model}Theoretical Framework}

We simulate nucleon-induced fission of uranium nuclei using a two-stage dynamical-statistical approach that couples the IQMD model with the GEMINI++ code event by event~\cite{Su2011PRC,Su2018PRC,Su2019,Song2022CPC}.
In the first stage, the IQMD model describes the early nonequilibrium evolution and preequilibrium particle emission.
In the second stage, the excited heavy residual nucleus formed at the end of the dynamical evolution is passed event by event to GEMINI++ for subsequent statistical decay~\cite{Charity2010,Mancusi2010PRC}.

\subsection{\label{sec_iqmd}Isospin-dependent quantum molecular dynamics model}

In the IQMD model, each nucleon $i$ in the colliding system is represented by a Gaussian wave packet of spatial width parameter $L$~\cite{Aichelin1991,Hartnack1998,Su2011PRC,Su2018PRC,Song2022CPC}:
\begin{equation}
\label{eq_wavepacket}
\begin{aligned}
\phi_i(\mathbf{r}, t) &= \frac{1}{(2\pi L)^{3/4}} \\
&\quad \times \exp\left[ -\frac{(\mathbf{r} - \mathbf{r}_i(t))^2}{4L} + \frac{i}{\hbar}\mathbf{r} \cdot \mathbf{p}_i(t) \right],
\end{aligned}
\end{equation}
where $\mathbf{r}_i(t)$ and $\mathbf{p}_i(t)$ denote the time-dependent mean position and momentum of the $i$th nucleon.
The $N_{\mathrm{tot}}$-body wave function is assumed to be the direct product of these single-nucleon wave packets:
\begin{equation}
\label{eq_manybody}
\Phi(\mathbf{x}_1, \dots, \mathbf{x}_{N_{\mathrm{tot}}}, t) = \prod_{i=1}^{N_{\mathrm{tot}}} \phi_i(\mathbf{x}_i, t),
\end{equation}
where $\mathbf{x}_i$ is the spatial coordinate of the $i$th nucleon and $N_{\mathrm{tot}}$ is the total number of nucleons.

The baryon Hamiltonian $H$ consists of the kinetic energy $T$, the Coulomb potential energy $U_{\mathrm{Coul}}$, and the nuclear potential energy:
\begin{equation}
\label{eq_hamiltonian}
H = T + U_{\mathrm{Coul}} + \int V(\rho, \delta)\,\mathrm{d}^3\mathbf{r},
\end{equation}
where the nuclear potential energy density $V(\rho, \delta)$ depends on the nucleon density $\rho$, its spatial gradient $\nabla\rho$, and the isospin asymmetry $\delta = (\rho_n - \rho_p)/\rho$:
\begin{equation}
\label{eq_potdensity}
\begin{aligned}
V(\rho, \delta) &= \frac{\alpha}{2}\frac{\rho^2}{\rho_0} + \frac{\beta}{\gamma + 1}\frac{\rho^{\gamma + 1}}{\rho_0^\gamma} + \frac{g_{\mathrm{sur}}}{2}(\nabla\rho)^2 \\
&\quad + \frac{C_{\mathrm{sp}}}{2}\left(\frac{\rho}{\rho_0}\right)^{\gamma_i}\rho\delta^2.
\end{aligned}
\end{equation}
Here $\rho_0 = 0.16~\mathrm{fm}^{-3}$ is the normal nuclear matter saturation density, $\rho_n$ and $\rho_p$ are the neutron and proton densities, and the local nuclear matter density is
\begin{equation}
\label{eq_density}
\rho(\mathbf{r}, t) = \frac{1}{(2\pi L)^{3/2}} \sum_{i=1}^{N_{\mathrm{tot}}} \exp\left[ -\frac{(\mathbf{r} - \mathbf{r}_i(t))^2}{2L} \right].
\end{equation}
In Eq.~(\ref{eq_potdensity}), the first two terms describe the two-body and three-body Skyrme interactions, while the third and fourth terms are the surface and symmetry potential energy densities.
The parameters are $\alpha = -356.00~\mathrm{MeV}$, $\beta = 303.00~\mathrm{MeV}$, $\gamma = 7/6$, $g_{\mathrm{sur}} = 130.00~\mathrm{MeV\,fm^5}$, $C_{\mathrm{sp}} = 38.06~\mathrm{MeV}$, and $\gamma_i = 0.75$~\cite{Hartnack1998,Su2018PRC,Song2022CPC}.
Under the self-consistently generated mean field, the time evolution of the baryons is governed by Hamilton's equations of motion:
\begin{equation}
\label{eq_hamilton}
\dot{\mathbf{r}}_i = \nabla_{\mathbf{p}_i} H, \qquad \dot{\mathbf{p}}_i = -\nabla_{\mathbf{r}_i} H.
\end{equation}

To simulate the short-range repulsive residual interaction and stochastic changes of the phase-space distribution, binary nucleon-nucleon ($NN$) collisions are included in the IQMD model.
The differential cross section of the $NN$ collisions is written as
\begin{equation}
\label{eq_nncross}
\left(\frac{\mathrm{d}\sigma}{\mathrm{d}\Omega}\right)_c = \sigma_c^{\mathrm{free}} f_c^{\mathrm{angl}} f_c^{\mathrm{med}},
\end{equation}
where $\sigma_c^{\mathrm{free}}$ is the free-space cross section, $f_c^{\mathrm{angl}}$ is the angular distribution factor, and $f_c^{\mathrm{med}}$ is the in-medium correction factor for channel $c$.
The subscript $c$ denotes elastic proton-proton, neutron-neutron, and neutron-proton scatterings as well as inelastic $NN$ collisions.
Above the pion-production threshold, inelastic collisions include the excitation and decay of $\Delta(1232)$ resonances via $NN \leftrightarrow N\Delta$ and $\Delta \leftrightarrow N\pi$.
The produced pions propagate under the Coulomb field of the system.
To enhance ground-state stability and preserve the fermionic nature of the $N_{\mathrm{tot}}$-body system, Pauli blocking and the phase-space density constraint method are applied.

During the IQMD evolution, prefragments are identified in phase space using a minimum spanning tree coalescence algorithm.
After the initial projectile-target interaction stage, following the excitation-energy stopping criterion~\cite{Su2018PRC,Su2019,Song2022CPC}, the IQMD simulation terminates once the excitation energy per nucleon of the heaviest prefragment falls below $3~\mathrm{MeV}$.
Nucleons and light clusters separated in phase space are then recorded as preequilibrium particles, while the heavy residual nucleus is characterized by its charge number $Z$, mass number $A$, momentum, and center-of-mass excitation energy
\begin{equation}
\label{eq_iqmd_ex}
E^* = T_{\mathrm{res}}^{\mathrm{c.m.}} + U_{\mathrm{res}} - E_{\mathrm{g.s.}}(Z, A),
\end{equation}
where $T_{\mathrm{res}}^{\mathrm{c.m.}}$ and $U_{\mathrm{res}}$ are its internal kinetic and total potential energies, and $E_{\mathrm{g.s.}}(Z, A)$ is the experimental ground-state energy.

\subsection{\label{sec_gemini}GEMINI++ code}

Each heavy residual nucleus with $(Z, A, E^*)$ from the IQMD stage enters GEMINI++ and deexcites through a sequence of binary decays until particle emission becomes energetically forbidden or is outcompeted by $\gamma$-ray emission~\cite{Charity2010,Mancusi2010PRC}.
At each binary decay step before the saddle point, the evaporation of light particles $i = n, p, d, t, {}^3\mathrm{He}$, and $\alpha$ competes directly with fission according to their partial decay widths.
For a parent nucleus with excitation energy $E^*$ and spin $J_m$, the partial decay width $\Gamma_i(E^*, J_m)$ for evaporating a light particle $i$ is calculated from the Hauser--Feshbach formalism:
\begin{equation}
\label{eq_gamma_evap}
\begin{aligned}
\Gamma_i(E^*, J_m) &= \frac{1}{2\pi\rho_m(u_0, J_m)} \sum_{J_d} \int_{0}^{E^* - B_i - E_{\mathrm{rot}}(J_d)} \\
&\quad \times \sum_{j = |J_m - J_d|}^{J_m + J_d} \sum_{\ell = |j - S_i|}^{j + S_i} T_\ell(\varepsilon)\,\rho_d(u_d, J_d)\,\mathrm{d}\varepsilon,
\end{aligned}
\end{equation}
where $B_i$, $S_i$, $\varepsilon$, $\ell$, and $j$ are the separation energy, intrinsic spin, kinetic energy, orbital angular momentum, and total angular momentum of the emitted particle $i$.
Here $J_d$ denotes the spin of the daughter nucleus, and $T_\ell(\varepsilon)$ is the transmission coefficient.
The parent and daughter level densities $\rho_m(u_0, J_m)$ and $\rho_d(u_d, J_d)$ are evaluated at thermal excitation energies $u_0 = E^* - E_{\mathrm{rot}}(J_m)$ and $u_d = E^* - B_i - E_{\mathrm{rot}}(J_d) - \varepsilon$, where $E_{\mathrm{rot}}(J)$ is the ground-state rotational energy.
The competing total fission decay width $\Gamma_f(E^*, J_m)$ is evaluated from the Bohr--Wheeler formalism:
\begin{equation}
\label{eq_gamma_fiss}
\begin{aligned}
&\Gamma_f(E^*, J_m) = \frac{1}{2\pi\rho_m(u_0, J_m)} \\
&\quad \times \int_{0}^{u_0 - B_f(J_m)} \rho_{\mathrm{sad}}\!\left(u_0 - B_f(J_m) - \varepsilon_f, J_m\right)\,\mathrm{d}\varepsilon_f,
\end{aligned}
\end{equation}
where $\rho_{\mathrm{sad}}$ is the level density at the fission saddle point, $\varepsilon_f$ is the kinetic energy in the fission degree of freedom at the saddle point, and $B_f(J_m)$ is the spin-dependent fission barrier.
Here $B_f(J_m)$ is taken from Sierk's finite-range liquid-drop model plus a constant barrier shift $\Delta B_f$, and no presaddle transient delay or intermediate-mass-fragment emission is included.
At each step, fission, $\gamma$-ray emission, or evaporation of particle $i$ is selected stochastically according to its partial width relative to $\Gamma_f + \Gamma_\gamma + \sum_i \Gamma_i$, where $\Gamma_\gamma$ is the $\gamma$-ray decay width.
Once a nucleus enters the fission channel, additional light particles can evaporate during the saddle-to-scission descent~\cite{Mancusi2010PRC}.
At scission, the primary-fragment mass distribution and total kinetic energy follow the Rusanov liquid-drop systematics~\cite{Rusanov1997,Itkis1998}, and the excited primary fragments subsequently decay via Eq.~(\ref{eq_gamma_evap}).

Because $\rho_m(u_0, J_m)$ cancels in the presaddle branching ratios, the competition in Eqs.~(\ref{eq_gamma_evap}) and~(\ref{eq_gamma_fiss}) is governed by $\rho_d$ and $\rho_{\mathrm{sad}}$.
By default, GEMINI++ uses an all-energy Fermi-gas (FG) level density calibrated on preactinide reactions~\cite{Charity2010,Mancusi2010PRC}.
To improve the low-energy level density in actinide fission, we implement a composite Gilbert--Cameron (GC) prescription for both $\rho_d$ and $\rho_{\mathrm{sad}}$~\cite{GilbertCameron1965,Koning2008}:
\begin{equation}
\label{eq_leveldensity}
\rho_{\mathrm{GC}}(u, J) =
\begin{cases}
T_m^{-1}\exp\{[u - E_0(J)]/T_m\}, & 0 < u \le u_m, \\[4pt]
\rho_{\mathrm{FG}}(u, J), & u > u_m,
\end{cases}
\end{equation}
where $u$ denotes the thermal excitation energy after subtracting the rotational energy, while pairing and shell shifts are incorporated within $\rho_{\mathrm{FG}}(u, J)$.
The matching energy is $u_m = (2.33 + 253/A)~\mathrm{MeV} + \Delta_p$, with $\Delta_p = 24/\sqrt{A}$, $12/\sqrt{A}$, and $0~\mathrm{MeV}$ for even-even, odd-$A$, and odd-odd daughter nuclei and $\Delta_p = 0$ at the saddle point.
At $u = u_m$, $T_m$ is set to the Fermi-gas nuclear temperature and $E_0(J)$ is determined for each spin $J$ by the continuity of $\rho_{\mathrm{GC}}(u, J)$.
In $\rho_{\mathrm{sad}}$, the asymptotic level-density parameter is scaled relative to the ground-state parameter $a_n$ by the ratio $a_f/a_n$, with the actinide parameter choices benchmarked in Sec.~\ref{sec_stages}.

\section{\label{sec_results}Results and Discussion}

\subsection{\label{sec_stages}Emission stages and residual nuclei}

\begin{figure*}[!t]
\centering
\includegraphics[width=0.88\textwidth]{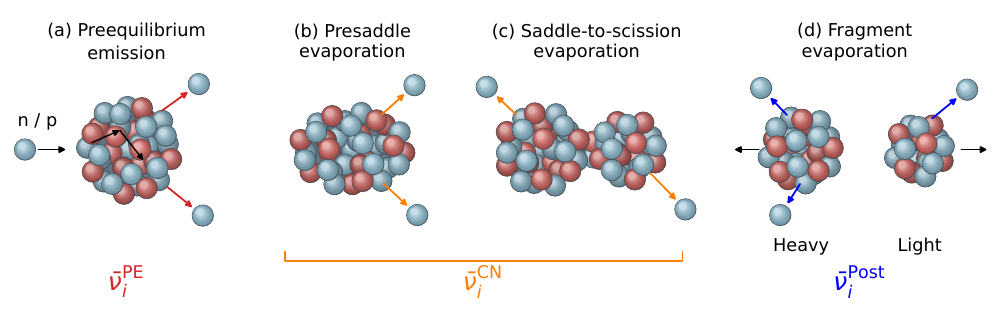}
\caption{\label{fig:fission_process}Schematic of the three emission stages: (a) PE emission in IQMD, (b) presaddle and (c) saddle-to-scission evaporation in GEMINI++, which together form the CN component, and (d) Post evaporation from the primary fragments.}
\end{figure*}

\begin{figure}[!htbp]
\centering
\includegraphics[width=\columnwidth]{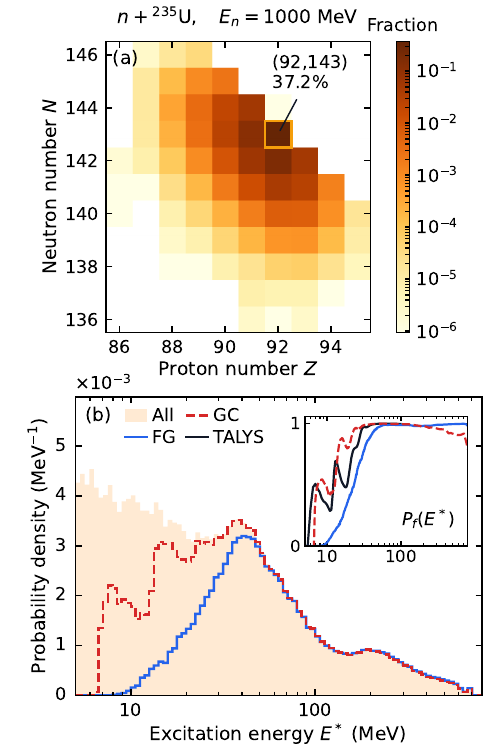}
\caption{\label{fig:iqmd_transition}Residual nuclei with $Z \ge 80$ and $A \ge 200$ from IQMD in $n + {}^{235}\mathrm{U}$ at $E_n = 1000~\mathrm{MeV}$: (a) distribution in the $(Z, N)$ plane and (b) $E^*$ distribution of $^{235}\mathrm{U}$ residual nuclei above $5~\mathrm{MeV}$ (shaded histogram) and the fissioning contributions under the FG (blue solid line) and GC (red dashed line) settings, normalized to all $^{235}\mathrm{U}$ residual nuclei per $\mathrm{MeV}$. The inset compares $P_f(E^*)$ with TALYS (black solid line)~\cite{Koning2023}.}
\end{figure}

Figure~\ref{fig:fission_process} classifies light-particle emission in fission events into three stages, using neutrons as an example.
Particles emitted during the IQMD stage define the preequilibrium (PE) component in Fig.~\ref{fig:fission_process}(a).
Particles evaporated before scission in GEMINI++, covering the presaddle chain and the saddle-to-scission descent in Figs.~\ref{fig:fission_process}(b) and~\ref{fig:fission_process}(c), form the compound-nucleus (CN) component (where the decaying nucleus after preequilibrium emission is a prefragment rather than a true compound nucleus, with ``CN'' retained by convention).
Particles evaporated from the primary fragments in Fig.~\ref{fig:fission_process}(d) form the postscission (Post) component.
For each species $i = n, p, d, t, {}^3\mathrm{He}$, and $\alpha$, the average multiplicity per fission event satisfies
\begin{equation}
\label{eq_nu_decomp}
\bar{\nu}_i = \bar{\nu}_i^{\mathrm{PE}} + \bar{\nu}_i^{\mathrm{CN}} + \bar{\nu}_i^{\mathrm{Post}},
\end{equation}
where $\bar{\nu}_i$ and its stage components denote the ensemble averages over fission events of the single-event multiplicities $\nu_i$, $\nu_i^{\mathrm{PE}}$, $\nu_i^{\mathrm{CN}}$, and $\nu_i^{\mathrm{Post}}$.
Stage multiplicities and probability distributions include fission events without emission, so the probabilities of integer multiplicities in each stage sum to unity.
In the calculations below, $10^5$ IQMD events are simulated at each incident energy for each reaction, whereas a larger event sample is used for the $1000~\mathrm{MeV}$ residual-nucleus benchmark in Fig.~\ref{fig:iqmd_transition}.

In the two-stage framework, each average multiplicity in Eq.~(\ref{eq_nu_decomp}) is conditioned on fission across the ensemble of residual nuclei at incident energy $E_{\mathrm{inc}}$:
\begin{equation}
\label{eq_fiss_conv}
\bar{\nu}_i(E_{\mathrm{inc}}) = \frac{\sum_{Z,N}\!\!\int w_{\mathrm{res}}(Z,N,E^*) P_f(Z,N,E^*) \langle \nu_i \rangle_f\,\mathrm{d}E^*}{\sum_{Z,N}\!\!\int w_{\mathrm{res}}(Z,N,E^*) P_f(Z,N,E^*)\,\mathrm{d}E^*},
\end{equation}
where $w_{\mathrm{res}}(Z, N, E^*)$ is the yield distribution of residual nuclei from IQMD, $P_f(Z, N, E^*)$ is the fission probability in GEMINI++, and $\langle \nu_i \rangle_f$ is the mean multiplicity for fissioning $(Z, N, E^*)$ nuclei.
Figure~\ref{fig:iqmd_transition} illustrates the distribution $w_{\mathrm{res}}$ and fission weight $w_{\mathrm{res}}P_f$ in Eq.~(\ref{eq_fiss_conv}) for residual nuclei with $Z \ge 80$ and $A \ge 200$ in $n + {}^{235}\mathrm{U}$ reactions at $E_n = 1000~\mathrm{MeV}$.
In Fig.~\ref{fig:iqmd_transition}(a), $37.2\%$ of the residual nuclei retain the $^{235}\mathrm{U}$ target composition, whereas nucleon-emission, charge-exchange, and pion channels populate $Z = 86\text{--}95$ and $N = 136\text{--}146$.
Residual nuclei with $Z < 92$ result from $(n,p)$ charge exchange and preequilibrium proton or cluster emission.
Residual nuclei with $Z > 92$ or $N > 144$ arise from $NN \to N\Delta \to NN\pi^\mp$ channels in which an escaping charged pion shifts the nuclear charge~\cite{Mancusi2010PRC,Boudard2013}.
Figure~\ref{fig:iqmd_transition}(b) shows the $E^*$ distribution of $^{235}\mathrm{U}$ residual nuclei together with the fissioning contributions weighted by $P_f(E^*)$.
Above $5~\mathrm{MeV}$, the distribution exhibits a broad plateau up to $40~\mathrm{MeV}$, falls steeply across $40\text{--}140~\mathrm{MeV}$, and extends to about $700~\mathrm{MeV}$ with a weak shoulder near $200~\mathrm{MeV}$.

Figure~\ref{fig:iqmd_transition}(b) also compares a modified GC setting with $\Delta B_f = -1~\mathrm{MeV}$ and $a_f/a_n = 1.02$ against the default FG setting with $\Delta B_f = 0$ and $a_f/a_n = 1.036$~\cite{Mancusi2010PRC}.
In GEMINI++, $\Delta B_f$ mainly governs fission at low $E^*$ near threshold, whereas $a_f/a_n$ controls the competition with particle evaporation at high $E^*$~\cite{LoMeo2015}.
Across actinide nuclei, Sierk's macroscopic liquid-drop barrier omits saddle-point microscopic corrections of order $1~\mathrm{MeV}$.
Consequently, the default FG setting suppresses fission at low $E^*$ and underestimates $P_f(E^*)$ below $40~\mathrm{MeV}$ compared with the $^{235}\mathrm{U}^*$ compound-nucleus benchmark from TALYS~\cite{Koning2023}.
Combining $\Delta B_f = -1~\mathrm{MeV}$ with the GC matching removes this low-energy suppression and reproduces the multichance step structure of TALYS.
At high $E^*$, diffuse-surface Fermi-gas theory relates $a_f/a_n - 1$ to the saddle-point surface-area increase, which is smaller in high-fissility actinides with compact macroscopic saddle shapes than in preactinides~\cite{Toke1981}.
Spallation analyses without presaddle transient delay also show that retaining the preactinide value $a_f/a_n = 1.036$ overestimates uranium fission probabilities at high $E^*$ and call for smaller effective ratios of $1.00\text{--}1.02$~\cite{Tishchenko2005}.
Adopting $a_f/a_n = 1.02$ moderates $P_f(E^*)$ above $200~\mathrm{MeV}$, while fission events at high $E^*$ cool before scission mainly through evaporation during the saddle-to-scission descent.
Because over $99\%$ of the residual nuclei in Fig.~\ref{fig:iqmd_transition}(a) are actinides with $Z \ge 89$ sharing these barrier and saddle-shape properties, this modified GC setting is applied uniformly across the ensemble of residual nuclei.

\begin{figure}[!htbp]
\centering
\includegraphics[width=\columnwidth]{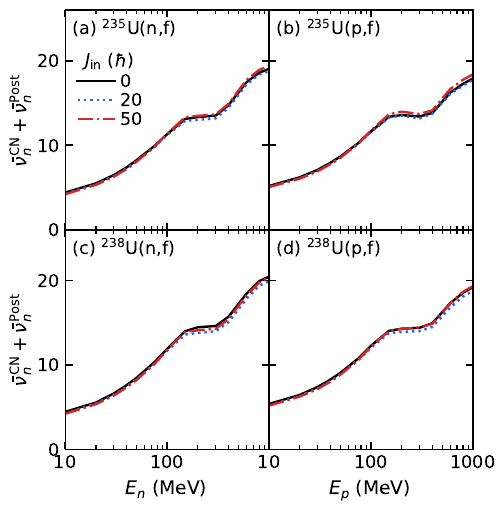}
\caption{\label{fig:spin}Statistical evaporation neutron multiplicity $\bar{\nu}_n^{\mathrm{CN}} + \bar{\nu}_n^{\mathrm{Post}}$ as a function of incident energy for (a) $^{235}\mathrm{U}(n,f)$, (b) $^{235}\mathrm{U}(p,f)$, (c) $^{238}\mathrm{U}(n,f)$, and (d) $^{238}\mathrm{U}(p,f)$. The black solid, blue dotted, and red dash-dotted lines denote calculations with initial spins $J_{\mathrm{in}} = 0$, $20\hbar$, and $50\hbar$ for even-$A$ nuclei, increased by $\hbar/2$ for odd-$A$ nuclei.}
\end{figure}

Because the semiclassical IQMD model does not provide quantized spins, the calculations pass residual nuclei to GEMINI++ with initial spin $J_{\mathrm{in}} = 0$ for even $A$ and $\hbar/2$ for odd $A$.
Figure~\ref{fig:spin} tests this setting by comparing $\bar{\nu}_n^{\mathrm{CN}} + \bar{\nu}_n^{\mathrm{Post}}$ for $J_{\mathrm{in}} = 0$, $20\hbar$, and $50\hbar$ in even-$A$ nuclei, corresponding to $\hbar/2$, $41\hbar/2$, and $101\hbar/2$ in odd-$A$ nuclei.
Across all four reactions, the three curves overlap closely from $10$ to $1000~\mathrm{MeV}$.
Below $50~\mathrm{MeV}$, a larger $J_{\mathrm{in}}$ lowers $B_f(J_m)$ in Eq.~(\ref{eq_gamma_fiss}), so fission occurs at an earlier chance with fewer presaddle neutrons while leaving more excitation energy for the fragments.
The resulting decrease in $\bar{\nu}_n^{\mathrm{CN}}$ and increase in $\bar{\nu}_n^{\mathrm{Post}}$, each within $0.6$ neutrons, nearly cancel.
Above $100~\mathrm{MeV}$, thermal excitation dominates over rotational energy, leaving the statistical neutron yields insensitive to $J_{\mathrm{in}}$.

\subsection{\label{sec_neutron}Neutron emission}

\begin{figure}[!htbp]
\centering
\includegraphics[width=\columnwidth]{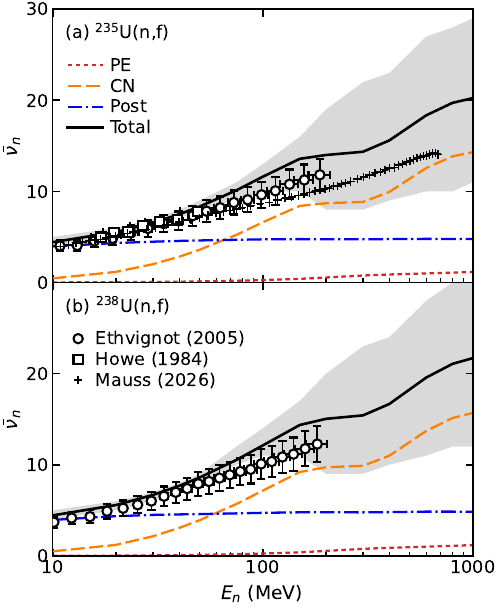}
\caption{\label{fig:nf_n_mult}Average neutron multiplicities as functions of $E_n$ for (a) $^{235}\mathrm{U}(n,f)$ and (b) $^{238}\mathrm{U}(n,f)$. The black solid, red short-dashed, orange long-dashed, and blue dash-dotted lines denote the calculated $\bar{\nu}_n$, $\bar{\nu}_n^{\mathrm{PE}}$, $\bar{\nu}_n^{\mathrm{CN}}$, and $\bar{\nu}_n^{\mathrm{Post}}$, and the gray shaded bands span the 16th to 84th percentiles, i.e., the central $68\%$ interval, of the multiplicity $\nu_n$. The open circles, open squares, and crosses represent the $\bar{\nu}_n$ data~\cite{Ethvignot2005,Howe1984,Mauss2026}.}
\end{figure}

Figure~\ref{fig:nf_n_mult} decomposes the calculated neutron multiplicities in $^{235,238}\mathrm{U}(n,f)$ into the three emission stages and compares $\bar{\nu}_n$ with the data~\cite{Howe1984,Ethvignot2005,Mauss2026}.
The measured total multiplicity $\bar{\nu}_n$ rises monotonically from about $4$ at $10~\mathrm{MeV}$ to $12.3$ near $180~\mathrm{MeV}$ in $^{238}\mathrm{U}(n,f)$ and about $14$ near $700~\mathrm{MeV}$ in $^{235}\mathrm{U}(n,f)$.
Below $40~\mathrm{MeV}$, the calculations stay within $0.8$ neutrons of the data and follow the measured rise across both reactions.
Above $100~\mathrm{MeV}$, the calculated $\bar{\nu}_n$ overestimates the data by $2.1\text{--}4.9$ neutrons because weak preequilibrium emission in IQMD leaves higher excitation energy in the residual nuclei.
Between $150$ and $300~\mathrm{MeV}$, the calculated $\bar{\nu}_n$ flattens into a plateau before rising steeply again above $300~\mathrm{MeV}$, whereas the measured $\bar{\nu}_n$ increases smoothly.
Below the pion-production threshold, the decreasing $NN$ elastic cross section and easier preequilibrium nucleon escape limit additional energy deposition into the residual nuclei across this interval.
Above $300~\mathrm{MeV}$, inelastic $NN \to N\Delta$ excitation and $\Delta$ or pion reabsorption deposit substantially more energy into the residual nuclei and restore the rapid growth of $\bar{\nu}_n$.
In the decomposition of Eq.~(\ref{eq_nu_decomp}), $\bar{\nu}_n^{\mathrm{Post}}$ increases only from $4.0$ at $10~\mathrm{MeV}$ to about $4.8$ above $100~\mathrm{MeV}$.
By contrast, $\bar{\nu}_n^{\mathrm{CN}}$ rises from below $0.5$, overtakes $\bar{\nu}_n^{\mathrm{Post}}$ at $60\text{--}70~\mathrm{MeV}$, and reaches $14.3$ in $^{235}\mathrm{U}(n,f)$ and $15.7$ in $^{238}\mathrm{U}(n,f)$ at $1000~\mathrm{MeV}$.
The larger $\bar{\nu}_n^{\mathrm{CN}}$ for $^{238}\mathrm{U}$ reflects lower neutron separation energies and higher fission barriers in its more neutron-rich residual nuclei.
From $10$ to $1000~\mathrm{MeV}$, this $\bar{\nu}_n^{\mathrm{CN}}$ increase of $13.9\text{--}15.2$ accounts for $88\%$ of the calculated $\bar{\nu}_n$ growth.
Meanwhile, $\bar{\nu}_n^{\mathrm{PE}}$ remains below $0.15$ up to $50~\mathrm{MeV}$ and reaches only $1.2$ at $1000~\mathrm{MeV}$.
Because neutron emission faces no Coulomb barrier, neutron evaporation dominates the prescission deexcitation of highly excited heavy residual nuclei~\cite{Rubchenya2007PRC,Mancusi2010PRC}.

\begin{figure}[!htbp]
\centering
\includegraphics[width=\columnwidth]{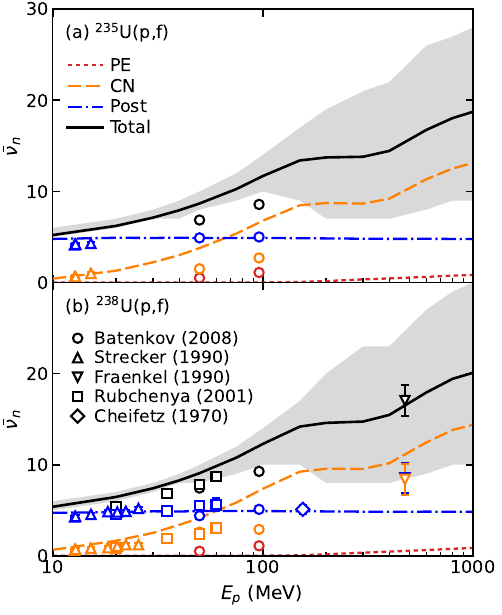}
\caption{\label{fig:pf_n_mult}Same as Fig.~\ref{fig:nf_n_mult}, but for (a) $^{235}\mathrm{U}(p,f)$ and (b) $^{238}\mathrm{U}(p,f)$ as functions of $E_p$. With the same color code as the calculated curves, the open circles, open upward and downward triangles, open squares, and open diamonds represent the data~\cite{Batenkov2008,Strecker1990,Fraenkel1990,Rubchenya2001,Cheifetz1970}.}
\end{figure}

Figure~\ref{fig:pf_n_mult} presents the stage-resolved neutron multiplicities for $^{235,238}\mathrm{U}(p,f)$.
In the coincidence data, kinematic analyses separate prescission and postscission yields~\cite{Strecker1990,Fraenkel1990,Rubchenya2001}, determine the postscission yield at $155~\mathrm{MeV}$~\cite{Cheifetz1970}, or resolve the PE, equilibrium, and Post components at $50$ and $96~\mathrm{MeV}$~\cite{Batenkov2008}, with the prescission and equilibrium yields plotted here against $\bar{\nu}_n^{\mathrm{CN}}$.
Across $E_p = 12.7\text{--}155~\mathrm{MeV}$, the postscission yield stays within $4.1\text{--}5.6$ and the PE yield reaches only $0.5\text{--}1.1$ at $50\text{--}96~\mathrm{MeV}$, whereas the prescission and equilibrium yields rise from $0.6$ to $2.7\text{--}3.1$.
At $475~\mathrm{MeV}$ in $^{238}\mathrm{U}(p,f)$, the prescission and postscission yields reach $8.4 \pm 1.7$ and $8.6 \pm 1.7$, raising the total yield from $5.4$ at $20~\mathrm{MeV}$ to $17.1 \pm 1.7$.
Thus, the data show that prescission evaporation drives the $\bar{\nu}_n$ growth below $155~\mathrm{MeV}$ and remains the largest contributor to this growth up to $475~\mathrm{MeV}$, although the single $475~\mathrm{MeV}$ point also gives a higher postscission yield.

The calculations in Fig.~\ref{fig:pf_n_mult} reproduce this stage-resolved trend across both reactions, with $\bar{\nu}_n^{\mathrm{Post}}$ staying near $4.8\text{--}4.9$ and $\bar{\nu}_n^{\mathrm{PE}}$ rising only to $0.8\text{--}0.9$ at $1000~\mathrm{MeV}$.
Meanwhile, $\bar{\nu}_n^{\mathrm{CN}}$ surpasses $\bar{\nu}_n^{\mathrm{Post}}$ above $E_p \approx 60~\mathrm{MeV}$ and reaches $13.1$ in $^{235}\mathrm{U}(p,f)$ and $14.4$ in $^{238}\mathrm{U}(p,f)$ at $1000~\mathrm{MeV}$, dominating the $\bar{\nu}_n$ growth.
Quantitatively, at $E_p = 12.7\text{--}15.2~\mathrm{MeV}$, $\bar{\nu}_n^{\mathrm{Post}}$ exceeds the data by $0.2\text{--}0.7$ neutrons, whereas at $18\text{--}35~\mathrm{MeV}$ in $^{238}\mathrm{U}(p,f)$, $\bar{\nu}_n^{\mathrm{CN}}$ and $\bar{\nu}_n$ exceed the data by $0.6\text{--}1.1$ neutrons~\cite{Strecker1990,Rubchenya2001}.
These low-energy differences reflect the omission of near-barrier quantum and shell effects in the semiclassical IQMD stage and in the liquid-drop scission systematics~\cite{Rusanov1997,Itkis1998}.
At $E_p = 50\text{--}96~\mathrm{MeV}$, $\bar{\nu}_n^{\mathrm{CN}}$ and $\bar{\nu}_n$ exceed the data by $1.6\text{--}4.2$ and $1.2\text{--}2.9$ neutrons~\cite{Rubchenya2001,Batenkov2008}, whereas $\bar{\nu}_n^{\mathrm{PE}} \approx 0.01$ underestimates the $0.5\text{--}1.1$ preequilibrium data~\cite{Batenkov2008}.
The weaker preequilibrium emission in IQMD leaves higher excitation energy in the residual nuclei, raising both $\bar{\nu}_n^{\mathrm{CN}}$ and $\bar{\nu}_n$.
At $E_p = 475~\mathrm{MeV}$ in $^{238}\mathrm{U}(p,f)$, the calculated $\bar{\nu}_n$ agrees with the total datum, whereas the data place both the prescission and postscission yields near $8.5$, below $\bar{\nu}_n^{\mathrm{CN}}$ and above $\bar{\nu}_n^{\mathrm{Post}}$~\cite{Fraenkel1990}.
Because agreement in $\bar{\nu}_n$ alone cannot independently validate the stage partitioning, further coincidence measurements are needed to constrain the source decomposition above $200~\mathrm{MeV}$.

\begin{figure}[!htbp]
\centering
\includegraphics[width=\columnwidth]{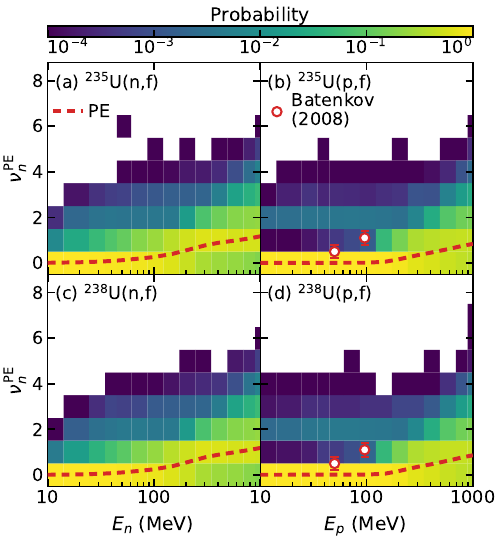}
\caption{\label{fig:pe_n_dist}Probability distributions of $\nu_n^{\mathrm{PE}}$ as functions of incident energy for (a) $^{235}\mathrm{U}(n,f)$, (b) $^{235}\mathrm{U}(p,f)$, (c) $^{238}\mathrm{U}(n,f)$, and (d) $^{238}\mathrm{U}(p,f)$. The color scale indicates the emission probability per fission event, the red short-dashed lines denote the calculated $\bar{\nu}_n^{\mathrm{PE}}$, and the red open circles in (b) and (d) represent the $\bar{\nu}_n^{\mathrm{PE}}$ data~\cite{Batenkov2008}. Cells near the $10^{-4}$ floor correspond to only a few fission events.}
\end{figure}

Beyond the averages, the gray shaded bands in Figs.~\ref{fig:nf_n_mult} and~\ref{fig:pf_n_mult}, spanning the 16th to 84th percentiles (central $68\%$) of $\nu_n$, widen from about one neutron at $10~\mathrm{MeV}$ to nearly $20$ neutrons at $1000~\mathrm{MeV}$.
To examine the origin of these fluctuations, Fig.~\ref{fig:pe_n_dist} shows the probability distributions of $\nu_n^{\mathrm{PE}}$.
In Figs.~\ref{fig:pe_n_dist}(a)--\ref{fig:pe_n_dist}(d), the distribution is concentrated at $\nu_n^{\mathrm{PE}} = 0$ below $20~\mathrm{MeV}$ and broadens toward $\nu_n^{\mathrm{PE}} = 1\text{--}7$ with increasing energy.
Even at $50~\mathrm{MeV}$ where $\bar{\nu}_n^{\mathrm{PE}} < 0.15$, $P(\nu_n^{\mathrm{PE}} \ge 1)$ is about $11\%$ and $P(\nu_n^{\mathrm{PE}} \ge 2)$ is about $0.5\%$ in $^{235,238}\mathrm{U}(n,f)$, whereas $P(\nu_n^{\mathrm{PE}} \ge 1)$ in $^{235,238}\mathrm{U}(p,f)$ stays near $0.5\%$.
At $1000~\mathrm{MeV}$, $^{235,238}\mathrm{U}(n,f)$ reaches $\bar{\nu}_n^{\mathrm{PE}} = 1.2$, where $P(\nu_n^{\mathrm{PE}} = 0)$ is about $18\%$ and $P(\nu_n^{\mathrm{PE}} \ge 2)$ is about $29\%$.
In $^{235,238}\mathrm{U}(p,f)$ at $1000~\mathrm{MeV}$, $\bar{\nu}_n^{\mathrm{PE}} \approx 0.85$, where $P(\nu_n^{\mathrm{PE}} = 0)$ is about $38\%$ and $P(\nu_n^{\mathrm{PE}} \ge 2)$ is about $19\%$.
Thus, $(n,f)$ reactions yield higher $\nu_n^{\mathrm{PE}} \ge 1$ probabilities than $(p,f)$ reactions because the incident neutron itself can be re-emitted as a fast preequilibrium neutron.
These fluctuations in $\nu_n^{\mathrm{PE}}$ leave the residual nuclei with a broad spread of mass numbers $A$ and excitation energies $E^*$.
Because the subsequent evaporation and fission widths depend nonlinearly on $A$ and $E^*$, the full ensemble of residual nuclei must be passed to GEMINI++ rather than a single average compound nucleus~\cite{Mancusi2010PRC,Manna2025PLB}.

\begin{figure}[!htbp]
\centering
\includegraphics[width=\columnwidth]{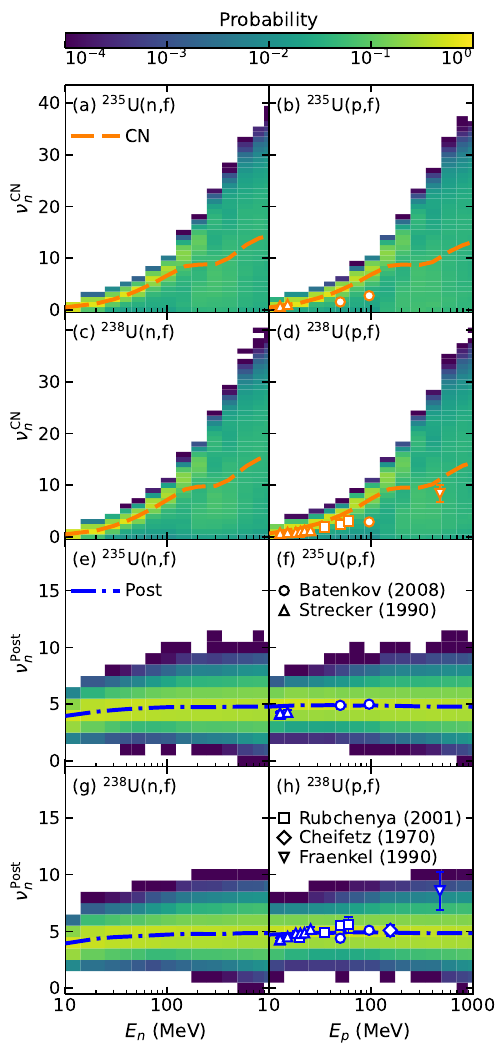}
\caption{\label{fig:cn_post_n_dist}Same as Fig.~\ref{fig:pe_n_dist}, but for (a)--(d) $\nu_n^{\mathrm{CN}}$ and (e)--(h) $\nu_n^{\mathrm{Post}}$. The orange long-dashed and blue dash-dotted lines denote the calculated $\bar{\nu}_n^{\mathrm{CN}}$ and $\bar{\nu}_n^{\mathrm{Post}}$, and the open circles, open upward triangles, open squares, open diamonds, and open downward triangles in (b), (d), (f), and (h) represent the corresponding data from Fig.~\ref{fig:pf_n_mult}~\cite{Batenkov2008,Strecker1990,Rubchenya2001,Cheifetz1970,Fraenkel1990}.}
\end{figure}

To track how this fluctuating ensemble of residual nuclei evolves during statistical decay, Fig.~\ref{fig:cn_post_n_dist} compares the probability distributions of $\nu_n^{\mathrm{CN}}$ and $\nu_n^{\mathrm{Post}}$.
In Figs.~\ref{fig:cn_post_n_dist}(a)--\ref{fig:cn_post_n_dist}(d), the $\nu_n^{\mathrm{CN}}$ distributions shift toward higher multiplicities and broaden rapidly with incident energy.
Specifically, $\nu_n^{\mathrm{CN}}$ is confined to $\nu_n^{\mathrm{CN}} \le 8$ at $50~\mathrm{MeV}$, whereas at $1000~\mathrm{MeV}$ the distribution extends across $\nu_n^{\mathrm{CN}} = 0\text{--}40$ with $\bar{\nu}_n^{\mathrm{CN}} = 13.1\text{--}15.7$.
Within $\bar{\nu}_n^{\mathrm{CN}}$, presaddle evaporation supplies nearly all yield at $10~\mathrm{MeV}$, becomes comparable to saddle-to-scission emission at intermediate energies, and contributes about $25\%$ at $1000~\mathrm{MeV}$ alongside $75\%$ from the saddle-to-scission descent.
By contrast, the $\nu_n^{\mathrm{Post}}$ distributions in Figs.~\ref{fig:cn_post_n_dist}(e)--\ref{fig:cn_post_n_dist}(h) remain narrow up to $1000~\mathrm{MeV}$, with $\bar{\nu}_n^{\mathrm{Post}} \approx 4.8$ and a central $68\%$ interval of $\nu_n^{\mathrm{Post}} = 4\text{--}6$.
The strong broadening of $\nu_n^{\mathrm{CN}}$ thus reflects the broad distribution of residual nuclei combined with both presaddle multichance paths and saddle-to-scission evaporation.

\begin{figure}[!htbp]
\centering
\includegraphics[width=\columnwidth]{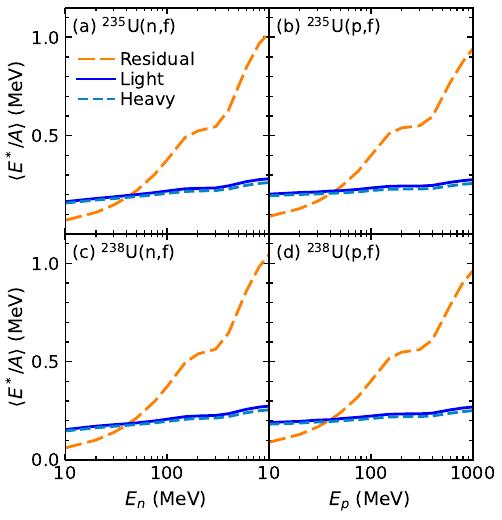}
\caption{\label{fig:excitation}Calculated $\langle E^*/A \rangle$ in fission events as a function of incident energy for (a) $^{235}\mathrm{U}(n,f)$, (b) $^{235}\mathrm{U}(p,f)$, (c) $^{238}\mathrm{U}(n,f)$, and (d) $^{238}\mathrm{U}(p,f)$. The orange long-dashed, blue solid, and cyan-blue short-dashed lines denote the residual nuclei and the primary light and heavy fragments.}
\end{figure}

To clarify the weak energy dependence of $\bar{\nu}_n^{\mathrm{Post}}$, Fig.~\ref{fig:excitation} compares the average excitation energy per nucleon $\langle E^*/A \rangle$ of the residual nuclei and primary fragments in fission events.
In Figs.~\ref{fig:excitation}(a)--\ref{fig:excitation}(d), $\langle E^*/A \rangle$ of the residual nuclei rises more than tenfold from below $0.1~\mathrm{MeV}$ at $10~\mathrm{MeV}$ to about $1.0~\mathrm{MeV}$ at $1000~\mathrm{MeV}$, whereas that of the primary fragments stays within $0.15\text{--}0.28~\mathrm{MeV}$ across $10\text{--}1000~\mathrm{MeV}$.
Below $40~\mathrm{MeV}$, the energy released in fission raises $\langle E^*/A \rangle$ of both fragments above that of the residual nuclei.
Above $40~\mathrm{MeV}$, prescission neutron evaporation removes most of the added excitation energy, widening the $\langle E^*/A \rangle$ gap between the residual nuclei and the fragments~\cite{Rubchenya2007PRC,Mancusi2010PRC}.
At scission, the remaining excitation energy and the fission energy release are partitioned between fragment total kinetic energy and the fragment excitation energies~\cite{Rusanov1997,Itkis1998}, limiting the thermal energy available for postscission evaporation.
In addition, $\langle E^*/A \rangle$ of the residual nuclei exhibits a weak shoulder across $150\text{--}300~\mathrm{MeV}$ before steepening above the $NN \to N\Delta$ threshold, directly mirroring the intermediate-energy plateau in Figs.~\ref{fig:nf_n_mult} and~\ref{fig:pf_n_mult}.
Thus, most of the added excitation energy of the residual nuclei is carried away by prescission evaporation rather than deposited into the fission fragments, explaining the weak energy dependence of $\bar{\nu}_n^{\mathrm{Post}}$.

\subsection{\label{sec_lcp}Light charged-particle emission}

\begin{figure}[!htbp]
\centering
\includegraphics[width=\columnwidth]{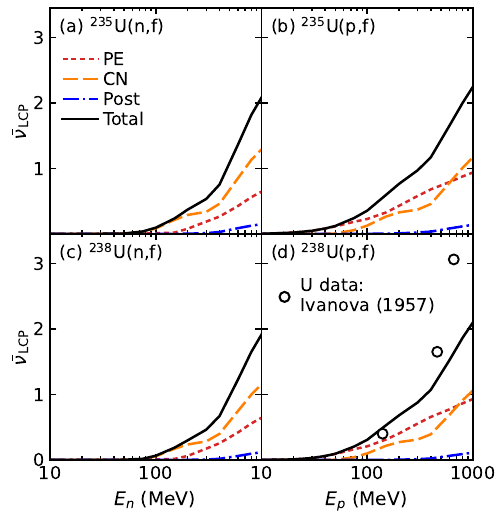}
\caption{\label{fig:charged_total}Average light charged-particle multiplicities as functions of incident energy for (a) $^{235}\mathrm{U}(n,f)$, (b) $^{235}\mathrm{U}(p,f)$, (c) $^{238}\mathrm{U}(n,f)$, and (d) $^{238}\mathrm{U}(p,f)$. The black solid, red short-dashed, orange long-dashed, and blue dash-dotted lines denote the calculated $\bar{\nu}_{\mathrm{LCP}}$, $\bar{\nu}_{\mathrm{LCP}}^{\mathrm{PE}}$, $\bar{\nu}_{\mathrm{LCP}}^{\mathrm{CN}}$, and $\bar{\nu}_{\mathrm{LCP}}^{\mathrm{Post}}$, and the open circles in (d) represent the $\bar{\nu}_{\mathrm{LCP}}$ data for natural uranium~\cite{Ivanova1957}.}
\end{figure}

Figure~\ref{fig:charged_total} shows the total light charged-particle (LCP) multiplicity $\bar{\nu}_{\mathrm{LCP}}$, summed over $p$, $d$, $t$, ${}^3\mathrm{He}$, and $\alpha$, together with its stage components.
Because of the Coulomb barrier, $\bar{\nu}_{\mathrm{LCP}}$ remains about an order of magnitude below $\bar{\nu}_n$~\cite{Charity2010,Herbach2006NPA}.
In Fig.~\ref{fig:charged_total}(d), the data for natural uranium, which contains $99.3\%$ $^{238}\mathrm{U}$, increase from $0.4$ at $E_p = 140~\mathrm{MeV}$ to $1.65$ at $460~\mathrm{MeV}$ and $3.06$ at $660~\mathrm{MeV}$~\cite{Ivanova1957}.
The calculations agree with the data at $140~\mathrm{MeV}$ and follow the upward trend, while underestimating the data at $460\text{--}660~\mathrm{MeV}$.
This high-energy underestimation reflects weaker preequilibrium proton and phase-space cluster emission in IQMD above $400~\mathrm{MeV}$.
Across Figs.~\ref{fig:charged_total}(a)--\ref{fig:charged_total}(d), $\bar{\nu}_{\mathrm{LCP}}$ rises from near zero at low energies to $1.9\text{--}2.2$ at $1000~\mathrm{MeV}$, driven almost entirely by $\bar{\nu}_{\mathrm{LCP}}^{\mathrm{PE}}$ and $\bar{\nu}_{\mathrm{LCP}}^{\mathrm{CN}}$.
In $^{235,238}\mathrm{U}(n,f)$, $\bar{\nu}_{\mathrm{LCP}}$ stays below $0.01$ up to $50~\mathrm{MeV}$, whereas above $100~\mathrm{MeV}$ $\bar{\nu}_{\mathrm{LCP}}^{\mathrm{CN}}$ rises ahead of $\bar{\nu}_{\mathrm{LCP}}^{\mathrm{PE}}$ and reaches $1.1\text{--}1.3$ versus $0.6$ at $1000~\mathrm{MeV}$.
By contrast, in $^{235,238}\mathrm{U}(p,f)$, $\bar{\nu}_{\mathrm{LCP}}^{\mathrm{PE}}$ rises above $20~\mathrm{MeV}$ and exceeds $\bar{\nu}_{\mathrm{LCP}}^{\mathrm{CN}}$ up to $600~\mathrm{MeV}$, before $\bar{\nu}_{\mathrm{LCP}}^{\mathrm{CN}}$ overtakes it between $600$ and $800~\mathrm{MeV}$.
Because the primary fragments have low excitation energies in Fig.~\ref{fig:excitation}, the Coulomb barrier keeps $\bar{\nu}_{\mathrm{LCP}}^{\mathrm{Post}}$ below $0.15$ up to $1000~\mathrm{MeV}$.

\begin{figure}[!htbp]
\centering
\includegraphics[width=\columnwidth]{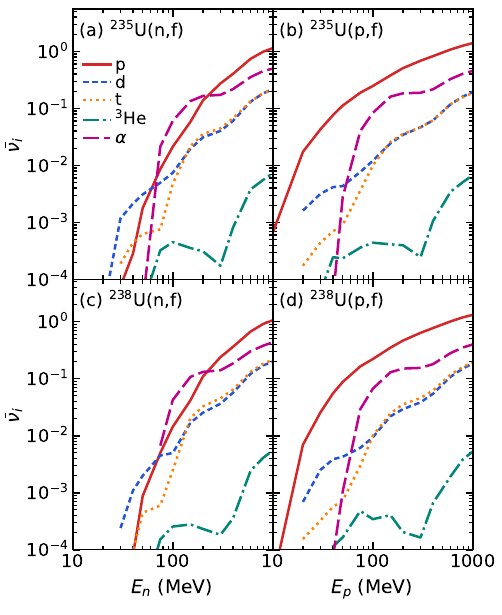}
\caption{\label{fig:charged_species}Average multiplicities $\bar{\nu}_i$ of individual light charged-particle species as functions of incident energy for (a) $^{235}\mathrm{U}(n,f)$, (b) $^{235}\mathrm{U}(p,f)$, (c) $^{238}\mathrm{U}(n,f)$, and (d) $^{238}\mathrm{U}(p,f)$. The red solid, blue short-dashed, orange dotted, green dash-dotted, and purple long-dashed lines denote the calculated $\bar{\nu}_p$, $\bar{\nu}_d$, $\bar{\nu}_t$, $\bar{\nu}_{{}^3\mathrm{He}}$, and $\bar{\nu}_\alpha$.}
\end{figure}

Figure~\ref{fig:charged_species} resolves $\bar{\nu}_{\mathrm{LCP}}$ into the average multiplicities $\bar{\nu}_i$ of individual species.
Above $250~\mathrm{MeV}$, all four reactions follow $\bar{\nu}_p > \bar{\nu}_\alpha > \bar{\nu}_d \approx \bar{\nu}_t > \bar{\nu}_{{}^3\mathrm{He}}$, consistent with the inclusive yield ordering measured in $1.2~\mathrm{GeV}$ proton-induced spallation of heavy nuclei up to thorium~\cite{Herbach2006NPA,Herbach2006NIMA}.
At $1000~\mathrm{MeV}$, the multiplicities reach $\bar{\nu}_p = 1.1\text{--}1.4$, $\bar{\nu}_\alpha \approx 0.45$, $\bar{\nu}_d \approx \bar{\nu}_t \approx 0.2$, and $\bar{\nu}_{{}^3\mathrm{He}} \approx 6 \times 10^{-3}$.
This ordering is governed by the competition among the Coulomb barrier, separation energy, and direct preequilibrium scattering.
Protons face only a unit-charge barrier and can escape directly in preequilibrium collisions, keeping $\bar{\nu}_p$ largest across all energies in $(p,f)$ and above $250~\mathrm{MeV}$ in $(n,f)$.
Among composite clusters, the $28.3~\mathrm{MeV}$ binding energy of the $\alpha$ particle lowers its separation energy in heavy nuclei~\cite{Charity2010,Mancusi2010PRC}.
Prescission CN evaporation supplies over $95\%$ of $\bar{\nu}_\alpha$, placing $\bar{\nu}_\alpha$ ahead of $\bar{\nu}_d$ and $\bar{\nu}_t$ above $70~\mathrm{MeV}$ and even above $\bar{\nu}_p$ at $E_n \approx 75\text{--}230~\mathrm{MeV}$ in $(n,f)$.
Deuterons and tritons are also emitted mainly in the CN stage above $100~\mathrm{MeV}$ with intermediate yields.
Below $300~\mathrm{MeV}$, the $Z = 2$ Coulomb barrier and large separation energy suppress statistical ${}^3\mathrm{He}$ evaporation, leaving only preequilibrium emission of order $10^{-4}$.
This preequilibrium yield decreases from $100$ to $300~\mathrm{MeV}$ as the falling $NN$ elastic cross section and larger recoil momenta hinder phase-space coalescence.
Above $300~\mathrm{MeV}$, where $\langle E^*/A \rangle$ of the residual nuclei rises steeply in Fig.~\ref{fig:excitation}, CN evaporation of ${}^3\mathrm{He}$ opens and increases $\bar{\nu}_{{}^3\mathrm{He}}$ by over an order of magnitude.

\begin{figure}[!htbp]
\centering
\includegraphics[width=\columnwidth]{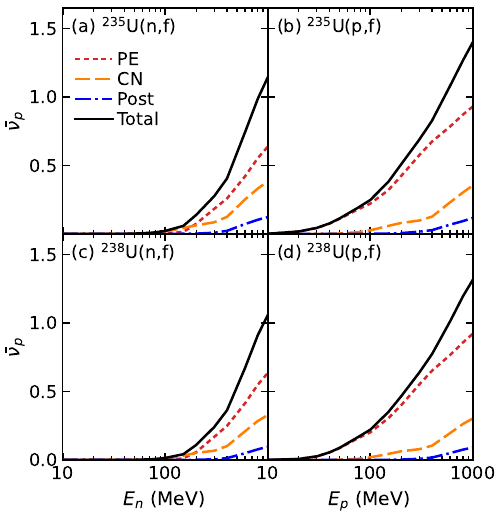}
\caption{\label{fig:proton_stages}Same as Fig.~\ref{fig:charged_total}, but for the calculated proton multiplicities $\bar{\nu}_p$, $\bar{\nu}_p^{\mathrm{PE}}$, $\bar{\nu}_p^{\mathrm{CN}}$, and $\bar{\nu}_p^{\mathrm{Post}}$.}
\end{figure}

Because protons dominate $\bar{\nu}_{\mathrm{LCP}}$ at high energies, Fig.~\ref{fig:proton_stages} presents the stage decomposition of $\bar{\nu}_p$.
Above $200~\mathrm{MeV}$, the proton components in all four reactions follow $\bar{\nu}_p^{\mathrm{PE}} > \bar{\nu}_p^{\mathrm{CN}} > \bar{\nu}_p^{\mathrm{Post}}$, in contrast to the CN-dominated neutron multiplicities in Figs.~\ref{fig:nf_n_mult} and~\ref{fig:pf_n_mult}.
At $1000~\mathrm{MeV}$, $\bar{\nu}_p^{\mathrm{PE}}$ reaches about $0.6$ in $(n,f)$ and $0.9$ in $(p,f)$.
These preequilibrium protons supply more than half of $\bar{\nu}_p$, consistent with the preequilibrium dominance observed in GeV spallation~\cite{Herbach2006NPA}, and account for nearly all of $\bar{\nu}_{\mathrm{LCP}}^{\mathrm{PE}}$ in Fig.~\ref{fig:charged_total}.
By contrast, Coulomb suppression limits statistical proton evaporation~\cite{Charity2010}, so $\bar{\nu}_p^{\mathrm{CN}}$ and $\bar{\nu}_p^{\mathrm{Post}}$ reach only $0.3\text{--}0.4$ and $0.1$ at $1000~\mathrm{MeV}$.
Thus, composite clusters account for about $70\%$ of $\bar{\nu}_{\mathrm{LCP}}^{\mathrm{CN}}$ in Fig.~\ref{fig:charged_total}.
In $(p,f)$, $\bar{\nu}_p^{\mathrm{PE}}$ rises above $E_p \approx 20~\mathrm{MeV}$ and yields $\bar{\nu}_p = 1.3\text{--}1.4$ at $1000~\mathrm{MeV}$.
In $(n,f)$, $\bar{\nu}_p^{\mathrm{PE}}$ becomes appreciable only around $E_n = 150\text{--}200~\mathrm{MeV}$, giving $\bar{\nu}_p \approx 1.1$ at $1000~\mathrm{MeV}$.
This difference arises because an incident proton can scatter and escape directly over the Coulomb barrier, whereas an incident neutron must transfer momentum to a target proton via neutron-proton collisions.

\begin{figure}[!htbp]
\centering
\includegraphics[width=\columnwidth]{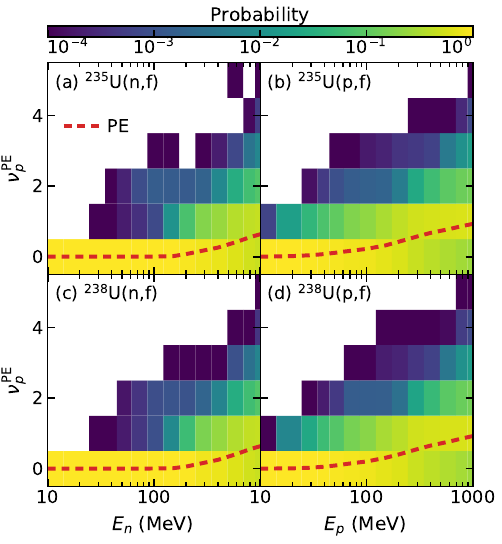}
\caption{\label{fig:pe_p_dist}Same as Fig.~\ref{fig:pe_n_dist}, but for $\nu_p^{\mathrm{PE}}$ and the calculated $\bar{\nu}_p^{\mathrm{PE}}$ (red short-dashed lines).}
\end{figure}

Figure~\ref{fig:pe_p_dist} shows the probability distributions of $\nu_p^{\mathrm{PE}}$.
In Figs.~\ref{fig:pe_p_dist}(a)--\ref{fig:pe_p_dist}(d), the distributions are concentrated at $\nu_p^{\mathrm{PE}} = 0$ below $20~\mathrm{MeV}$ and extend up to $\nu_p^{\mathrm{PE}} = 4\text{--}5$ at $1000~\mathrm{MeV}$, with nearly identical patterns for $^{235}\mathrm{U}$ and $^{238}\mathrm{U}$.
At $50~\mathrm{MeV}$, $P(\nu_p^{\mathrm{PE}} \ge 1)$ already reaches about $10\%$ in $(p,f)$, compared with less than $0.05\%$ in $(n,f)$.
At $1000~\mathrm{MeV}$, the emission probability $P(\nu_p^{\mathrm{PE}} \ge 1)$ reaches $75\%$ in $(p,f)$ compared with $54\%$ in $(n,f)$, dominated in both reactions by $\nu_p^{\mathrm{PE}} = 1$ events from direct proton escape.
Across both particle sectors, CN neutron evaporation, which faces no Coulomb barrier, removes most of the excess thermal energy before scission and stabilizes the primary-fragment excitation.
Meanwhile, preequilibrium proton emission and high-excitation CN cluster evaporation reduce the charge and fissility parameter $Z^2/A$ of the fissioning nuclei along the decay chain~\cite{Mancusi2010PRC}.

\section{\label{sec_summary}Summary}

Nucleon-induced fission at intermediate energies links fast nonequilibrium cascade dynamics to slow statistical evaporation and collective scission across an ensemble of residual nuclei.
Using the coupled IQMD--GEMINI++ framework, we have studied neutron and light charged-particle emission in $^{235,238}\mathrm{U}(n,f)$ and $^{235,238}\mathrm{U}(p,f)$ at incident energies from $10~\mathrm{MeV}$ to $1~\mathrm{GeV}$.
The calculated $\bar{\nu}_n$ follows the measured upward trend, with $\bar{\nu}_n^{\mathrm{CN}}$ providing the main contribution to the $\bar{\nu}_n$ increase across this energy range.
The $\nu_n^{\mathrm{CN}}$ distributions shift toward larger multiplicities and broaden rapidly with incident energy, whereas the $\nu_n^{\mathrm{Post}}$ distributions remain narrow near $\bar{\nu}_n^{\mathrm{Post}} \approx 4.8$.
Correspondingly, $\langle E^*/A \rangle$ rises steeply for the residual nuclei but increases only weakly for the primary fragments.
Above $40~\mathrm{MeV}$, prescission neutron evaporation removes most of the added excitation energy before scission, while the remaining energy is partitioned between fragment kinetic energy and the excitation of the primary fragments.

Despite small $\bar{\nu}_n^{\mathrm{PE}}$, the $\nu_n^{\mathrm{PE}}$ distributions exhibit multineutron tails at high energies and produce a broad distribution of residual nuclei in $A$ and $E^*$.
For light charged particles, $\bar{\nu}_{\mathrm{LCP}}$ is suppressed by the Coulomb barrier and follows $\bar{\nu}_p > \bar{\nu}_\alpha > \bar{\nu}_d \approx \bar{\nu}_t > \bar{\nu}_{{}^3\mathrm{He}}$ above $250~\mathrm{MeV}$.
At high energies, composite clusters are evaporated predominantly in the CN stage, whereas protons above $200~\mathrm{MeV}$ are emitted mainly in the PE stage with $\bar{\nu}_p^{\mathrm{PE}} > \bar{\nu}_p^{\mathrm{CN}} > \bar{\nu}_p^{\mathrm{Post}}$.
In the $\nu_p^{\mathrm{PE}}$ distributions, easier escape of the incident proton over the Coulomb barrier yields a higher preequilibrium proton emission probability in $(p,f)$ than in $(n,f)$, reaching $75\%$ versus $54\%$ at $1000~\mathrm{MeV}$.

Quantitative differences from the data remain in $\bar{\nu}_n$ above $100~\mathrm{MeV}$, in $(p,f)$ stage yields, and in high-energy light charged-particle multiplicities.
Combining the stage-resolved average multiplicities with their distributions shows that early emission shapes the composition and excitation energy of the residual nuclei, whereas prescission evaporation carries the growth of $\bar{\nu}_n$.
These results provide a reference for intermediate-energy fission modeling, where prescission particle emission alters the charge, mass, and excitation energy of the fissioning nuclei.

\begin{acknowledgments}
This work was supported by the National Natural Science Foundation of China under Grant No.~12475136.
\end{acknowledgments}

\section*{Data Availability}

The data that support the findings of this article are not publicly available.
The data are available from the authors upon reasonable request.

\bibliography{apssamp}

\end{document}